\documentclass[10pt,a4paper]{article}
\usepackage{amsmath}
\usepackage{amsthm}
\numberwithin{equation}{section}
\usepackage{graphicx}
\usepackage[T1]{fontenc}
\title{ Evidence for  Gross Domestic Product  growth  time delay dependence over  Foreign Direct Investment. A time-lag dependent correlation study.   }
	\author{    Marcel Ausloos$^{1,2,3,^*}$, Ali Eskandary$^{1}$, Parmjit Kaur$^4$,  Gurjeet Dhesi$5$
 }
\date{      $^1$  School of Business, University of Leicester,\\  Brookfield, Leicester, LE2 1RQ, United Kingdom\\Email: ma683@le.ac.uk 
  \vskip0.4cm $^2$ Department of Statistics and Econometrics, \\Bucharest University of Economic Studies, \\ Calea Dorobantilor 15-17, Bucharest, 010552 Sector 1, Romania.
  \vskip0.4cm $^3$  GRAPES, rue de la Belle Jardini\`ere, 483/0021\\ B-4031, Li\`ege Angleur, Belgium, Euroland \\Email: marcel.ausloos@ulg.ac.be  
  \vskip0.4cm 
$^4$  De Montfort University, Leicester Castle Business School, Hugh Aston, Leicester,  LE1 9BH,
 United Kingdom\\Email: pkcor@dmu.ac.uk
 \vskip0.4cm 
$^5$ School of Business, London South Bank University, 103, Borough Road, London, SE1 0AA, United Kingdom. 
\\Email: dhesig@lsbu.ac.uk
 \vskip0.8cm $^*$ corresponding author   
    }
\begin{document}
\maketitle

\begin{abstract}
This paper considers   an often forgotten   relationship, the time delay between a cause and its effect in economies and finance. We treat the case  of   Foreign Direct Investment  (FDI) and  economic growth, 
- measured through   a country  Gross Domestic Product (GDP).  The pertinent data refers to 43 countries, over 1970-2015, -  for a total of 4278 observations.
When countries are grouped 
 according to the Inequality-Adjusted Human Development Index (IHDI),   
it is found that a time lag dependence effect exists  in FDI-GDP  correlations.   
    This is established  through  a time-dependent Pearson 's product-moment correlation coefficient matrix.  
    Moreover, such a Pearson correlation coefficient  is observed  to evolve  from positive 
      to negative values depending on the  IHDI, from low to high.   It is   "politically and policy 
      "relevant" that 
      the correlation is statistically significant providing the time lag is less than 3 years.  A "rank-size" law is demonstrated.
      It is recommended to reconsider such a time lag  effect when discussing previous analyses whence conclusions on international business, and thereafter on forecasting. 
      \end{abstract}

 % \vskip0.2cm
   
  \vskip0.5cm
  
% Keywords: Foreign Direct Investment; Inequality-Adjusted Human Development Index; 
 %Pearson 's product-moment correlation coefficient  matrix; 
 %time lag econometrics; economic growth  (GDP)
 \vskip0.5cm
 
 %JEL Classification :  F21, F43, F35, C12, C22,  C53, 
% \newpage
\section{Introduction}

Foreign Direct Investment  (FDI)  concerns investments made by a company or individual  from one country into  business interests in another country, in order to establish  either effective control of, or at least substantial influence over, the decision making of the concerned  (foreign) business.  Literature has remained indecisive regarding the impact of  FDI upon the economic growth enjoyed by the nations in which some business is  targeted (Hoekman and  Javorcik,  2004).  The crux of this   debate  lies in the inability to reach a consensus on (necessarily asymmetric) causality (Javorcik and Saggi,  2010), but Barrell et al. (2017) observed the main factors affecting \underline{bilateral} FDI stocks  (from 14 high income countries to all OECD countries over the period 1995-2012). High market integration is a fundamental aspect (Barrell and Nahhas, 2018). One may expect the same at an "average level" measured through a country  Gross Domestic Product (GDP).

Our paper explores  such an area of research on  FDI through original measurements of time dependent econometrics effects. 

 %It seems indeed that one has forgotten to consider the  GDP time delay dependence from Foreign Direct Investment  toward stimulating growth in previous analyses.

Economies are measured in terms of  outputs  like the   GDP. This output is composed of the goods and services produced within the country in a given year. Economic "growth" is thus measured in the annual change in this GDP as compared to the previous year. The change between these two years  can be positive or negative.  
 Any references to growth in this paper refer to this annualized change in GDP (often measured in percents), after removing inflation.  

Most   observers detect a positive relationship between FDI and GDP,  but many counterarguments do exist.  Research undertaken by  Blomstr{\"o}m and Kokko (1996), de Mello (1999), Alfaro et al.  (2004),  for example, all observed a positive  correlation. However, Hymer (1960), Caves (1971), Lipsey and Sjoholm (2006), among others, found a negative relationship; see a more extensive list of disagreeing authors  below in Sect. \ref{LitRev}. 

Given   this disagreement,  
we seek to address one underlying assumption which requires clarification. We stress that there is a conceptual difference between causality and correlation (Liang, 2014; 2016).  The purpose of our study  is not to make any statements in relation to establishing means of causality between FDI and Growth, as in Borensztein et al. (1998),  
Nair-Reichert and Weinhold (2001), Zhang (2001),   Choe (2003), Kuo et al. (2003), Hoekman and  Javorcik (2004), Li and Liu (2005), De Gregorio (2005), Hansen and Rand (2006), Mah (2010) nor "strict local effects" (Brambilla et al., 2009; Hale and Long, 2011, Wang et al., 2013), for example. This is much discussed elsewhere as this non-exhaustive list of references may indicate. Rather, we assume  that the relationship  exists regardless of its strength or (microeconomic) cause, together with spatial correlations (Blonigen et al., 2007).  We   focus  on stressing that a seemingly hidden or even  immediately accepted hypothesis,  that is "time independent correlations",  may make the current body of research to be "less inconclusive".

Thus, the present   paper is written to 
 demonstrate  that the existence of a time delay regarding the influence  of FDI   inflows  upon stimulating economic growth, as measured by GDP growth, might be a key in resolving the discrepancies.  Besides,  one may wonder why such a time delay seems not to be taken into account.  Moreover, assuming that there is some relationship between FDI and GDP growth, we ask how long  it takes for this FDI growth to be reflected  in  GDP  growth.   To find such a time span, a so called   "memory relaxation time"  between inflow and output, in which a finite correlation value is found to be significant, seems of interest from a policy (and also from a political) point of view indeed.
 
"Finally",   it can be  intuitively  admitted that the macroeconomic forces behind  some cash inflow could influence a mature economy differently that  an underdeveloped  society. Thus, such a   categorization  seems also a  pertinent discussion. We use a  UN's  recently invented index, the Inequality-Adjusted Human Development Index (IHDI), devlopped from the Human Development Index (HDI)\footnote{$http://hdr.undp.org/en/content/inequality-adjusted-human-development-index-ihdi$}.  For completeness, let us recall that, quoting Harttgen and Klasen (2010), {\it 
"The HDI is a composite index that measures the average achievement in a country in three basic dimensions of human development: a long and healthy life, measured by life expectancy at birth; education, measured by the adult literacy rate and the gross school enrollment, and standard of living, measured by GDP per capita (UNDP, 2006)."}
Thus, 
the (initially invented)  Human Development Index (HDI)  considers only average achievements and does (or did) not take into account the distribution of human development within a country or by population sub-groups. Thereafter, the 
  IHDI is invented in order to combine a country’s average achievements in health, education and income with how those achievements are distributed among country’s population by “discounting” each dimension’s average value according to its level of inequality.
In brief, the difference between the IHDI and HDI is the human development cost of inequality.

If not noticed,  we do not search nor study a Granger causality scheme; one should warn the reader that correlation studies do not immediately uncover causal links. 

	\section{Literature Review: "the state  of the art".}\label{LitRev}
%  Due to the absence of literature on the existence of a time lag between  FDI and GDP, we    highlight the different perspectives on the equal time relationships between FDI and economic growth.  
  Causal  time lags have been seen and discussed  in many  other fields, of course,  but we will be short on such references for the sake of conciseness.  
However,  from a more general viewpoint, we should mention that an extensive FDI and economic growth literature review from 1994 to 2012 has been provided by 
  Almfraji and Almsafir (2014).
 
No need to elaborate on the notion that FDI are private capital flows from a parent firm to an entity based in a foreign country (Griffin and Pustay, 2007); Marin and Schnitzer, 2011).   One may consider that FDI can flow in two directions, outwards as an investor and inwards as an investor. In this paper, any reference to FDI will strictly refer to FDI inflows received by a country, usually that with the lowest Gross Domestic Product (GDP).   Economic growth is measured in the annual change in GDP (after removing inflation) as compared to the previous year. The change between these two years determines the growth or shrinkage of the GDP which is measured in percentages. Any references to growth in this paper refer to this yearly change in GDP.  

\subsection{%2.3 
 The relationship between Foreign Direct Investment and Gross Domestic Product Growth.}
Theoretically, FDI is believed to directly impact growth through increasing a countries capital stock and through  acting as a vehicle for the transfer of technology and knowledge which in turn creates new job opportunities.  For example, an
 interesting work about FDI analysis on regional level (Strat, 2014) can be quoted.
These transfers cause substantial economy-wide spillovers,  thus   boosting long term productivity economy-wide and not merely for the recipient firm (Rappaport, 2000, Mencinger 2003).  The net effect is an increase domestic productivity and acceleration in economic growth (Borensztein et al. 1998; de Mello, 1999; De Gregorio, 2005). With commercial bank lending increasingly scarce in the 1980's , many countries loosened FDI restrictions and offered aggressive subsidies and tax incentives in order to attract foreign capital inflows (World Bank, 1997). This led to a surge of private capital flows particularly toward businesses in developing economies.     

	\subsection{%2.4 
	On the positive relationship between Foreign Direct Investment and   Gross Domestic Product growth.}
Scholars taking the view that there is a positive relationship between FDI and Growth  
based their conviction on macroeconomic studies which use aggregate FDI flows (Abbes et al., 2015). 
 In establishing causality, 
  two distinct theories emerge; the "FDI-led growth hypothesis" and the "market size hypothesis".  The former believes  that FDI   stimulates growth through increasing a country 's capital stock and allowing for the transfer of technology and knowledge which in turn creates new job opportunities (De Gregorio, 2005). The latter believes that  GDP  growth gives the host country new investment opportunities which in turn lead  to a large FDI inflow (Mah, 2010).   Within this line many scholars seek to establish the dominant factors affecting the growth. Romer (1993) dubbed these factors as "idea gaps" between countries of different economic standing which FDI can help bridge. The dominant factors are found  to fall into four categories; labour force skills (de Mello, 1999; Ali et al., 2016), technology transfer (Boreinsztein et al., 1998;  de Mello, 1999; Hansen and Rand, 2006;  Tu and Tan,  2011), infrastructure and institutional development (Balusubramanyam et al., 1996; Hermes and Lensink, 2003; Durham, 2004), and trade liberalisation (Bengoa and Sachez-Roble, 2003).

Although these studies all find different factor determinants of the effectiveness of FDI, the results can be interpreted  from the hypotheses  that countries seek to improve their domestic financial systems and processes in order to create an environment which will allow FDI to thrive. It must be noted at once  that these macroeconomic studies do carry flaws. They do not control adequately for simultaneity bias, country-specific effects and   lagged dependent variables in growth regression.  We consider that these flaws  may create some bias in both the coefficient estimates and standard errors, suggesting that we might also duly criticize the "positive relationship" findings. 

\subsection{%2.5 
On the negative relationship  between Foreign Direct Investment and   Gross Domestic Product growth.}

A  fair body of literature challenges the  belief that FDI has a positive impact upon growth,   in view of  firm level microeconomic studies.  
	
One of the most renowned theories on the negative effect of FDI upon growth is raised by Hymer (1960) and Caves (1974) in the form of "control theory",   
  For example, Saltz (1992)   or  Huang (1998, 2003)  
 found a negative correlation  
 explained through monopolisation  
 which causes a lag in domestic demand.   
  Another reason as presented by Braunstein and Epstein (2002)   is the lack of local investments,  
  resulting in a reduction in tax revenues,  
with  driving down wages, - due to reduced competition.  
 Bos et al. (1974) even  found    
  an \underline{outflow} of profits which exceeded the level of new investment.  
 This is also a conclusion which can de drawn from G\"org and Greenaway (2002).  
	
	In view of explaining some discrepancy, Lipsey  (2002)  suggests that most of the evidence for wage spill overs is found in developed countries where there is wider access to knowledge, capital and technology and as such multinational entities   are less likely to prevent spill overs.  In fact, for emerging countries, 
Ahmed's (2012) study of Malaysia, for the 1999-2009 period,  
 inspected the influence of FDI on human capital, labour force, absorptive capacity and GDP. He found that FDI inflows contributed negatively to total productivity and thus economic growth.    This was mirrored by Mazenda's (2014) study on South Africa for the period 1960-2002.  
	
Thus, such  studies conclude  that the effect of FDI on GDP growth is mostly negative, and distinguish several ingredients, explaining some disagreement with the findings of the "positive relationship" authors.

\subsection{%2.6 
No relationship found  between Foreign Direct Investment and   Gross Domestic Product growth.}

Let  us be "complete": a  third camp exists which believes that  there is no conclusive evidence at all to establish that a correlation between FDI and Growth exists:
Germidis (1977), Mansfield and Romeo (1980), Haddad and Harrison (1993), Aitken and Harrison (1999),    Irandoust (2001), Louzi and Abadi (2011),  Carkovic and Levine (2005), Herzer et al.  (2008), Belloumi (2014), Aga (2014), Temiz and G\"okmen (2014),
   all failed to find any robust  conclusion on the existence of   growth stemming from FDI inflow, - using different data analysis techniques. However, no time lag was considered for the relationship process, whence our present concern about time dependent correlations.
 
 \subsection{%2.7 
 The study of time lags.}
%This seems a logical flaw. 
Of course, time dependence of correlations have been studied. In fact,  studies about time lags, in intuitively causal processes, have been undertaken in different financial, managerial, and accounting  topics.    A literature review of interest  on audit report lags is provided by Abernathy et al. (2017). 
A "delayed expected loss recognition and the risk profile of banks" was considered  by 
 Bushman and Williams (2015).  Li and  Mei (2012) analyzed the influences of a time delay on the stability of a market model with stochastic volatility.  
Other particular studies,  Miskiewicz (2012) and Ausloos and Lambiotte (2007),  suggest to examine the  key role of time lags, on "globalization", as also discussed by Cerqueti et al. (2018).
%whence these papers impact  on the formulation of this research paper.

Let us emphasize Miskiewicz (2012) who used a model of the stock market coupled with an economy to investigate the role of the time delay span on the information flow. An information flow was coupled into the stock market model which interacted with the economy. The observation made was that through the autocorrelation of absolute returns, cycles appeared as the time delays were increased. This meant that increasing delays in the information flow resulted in the increasing homogenization in the behaviors of actors in the stock market. This collectivization was found to relate closely to the price bubbles and crashes of real markets, explaining how bubbles and crashes form.  The study also found that after a certain time delay, further increasing this delay had no effect on the behavior of actors. %This is in fact similar to the research undertaken in this paper, where we look for trends (using the Pearson and Autocorrelation functions) and  whether an increase in information time delay is substituted by  a time delay in the economic growth statistics.   
 Interestingly,  Miskiewicz (2012) study discovered  a relaxation time of flow influence:  his model holds only until a certain time delay, - after which further delay of information is found to have no effect.  
	
Previously, Ausloos and Lambiotte (2007) studied correlations between the GDP of rich countries where GDP proxied a nation 's wealth. Yearly fluctuations of the GDP were calculated which were checked for correlations with the correlation measure (based on the Theil index). Time delays with the least correlations were removed in order to generate a structure  within  the network of countries. The study observed patterns being formed under the form of clusters in the countries network. This structure adhered to geography and fell in line with economic globalization which homogenized the economic development of countries. 

Our research  takes a similar approach to Ausloos and Lambiotte  (2007) and seeks to form a network based on the country's  wealth. The correlation measure is framed on an Index,  here the Inequality-Adjusted Human Development Index  (IHDI).   

Thereafter, the range of time lag effect, as discussed by Miskiewicz (2012), can be deduced, for  any "memory time span concern". 

%However,  we point out again that  there is  "surprisingly" no time lag research nor a fortiori model applied for FDI and GDP correlations.  
\section{%3.1
Data}
	
The dataset consists of the observation from 43 countries over the period from 1970-2015, as  obtained from the World Bank Databank\footnote{
 $www.databank.worldbank.org$}. The country selection depends on the completeness of the data. There are   4278 individual observations. The FDI data focused on net inflows over the aforementioned period normalized in current USD. The GDP data gathered was annual GDP growth. The Data was gathered subject to availability; countries were chosen based upon the Inequality-Adjusted Human Development Index (IHDI). This is a measurement of the living standards in a country adjusted for inequality. This data is made available by the United Nations Development Project and is compiled on an annual basis. The reasoning is to categorize developing nations effectively as to ensure fair comparisons given that vast variances in the distribution of wealth will affect the economic multiplier effects of FDI across the economy and as such in respects to GDP growth. The data is split into four panels which follow the UN's IHDI. These four panels are:
\begin{itemize}
\item
 Very High IHDI, which translates into an IHDI score between 1-0.80, the (13) economies falling within this bracket are Australia, Austria, Canada, Denmark, France, Finland, Germany, Netherlands, Sweden, Great Britain, Norway, Iceland and Ireland, called sub-sample S1 below;
\item
 High IHDI,  which translates into an IHDI score between 0.799- 0.70, the (11)  economies falling within this bracket are Argentina, Israel, Spain, Italy, United States of America, Portugal, Greece, Japan, Malta, Cyprus, Korea, called sub-sample S2 below;
\item
 Medium IHDI,  which translates into an IHDI score between 0.699-0.55, the (8) economies falling within this bracket are Uruguay, Sri Lanka,  %Trinidad and Tobago, 
Venezuela, Mexico, Peru, Mauritius, Chile and Turkey,  called sub-sample S3 below;
\item
 Low IHDI, which translates into an IHDI score of 0.549 and below, the  (11) economies falling within this bracket are the Philippines,  %Colombia, 
Paraguay, % Iran,
 Iraq, Bolivia, %Marocco, 
 South Africa, Nigeria, Niger, El Salvador, India, Nepal and Ghana,  called sub-sample S4 below
\end{itemize}
%The countries are distributed into  such groups  so that    each category  approximately contains the same number of countries.  Indeed, 
  Given the disparity between the number of Very High IHDI and Low IHDI countries, the data exhausted the availability of Very High IHDI and High IHDI countries, - but we selected a few Low IHDI countries at random,  in view of allowing an equivalent number of economies, in scale with the other  categories for some statistical coherence.   

Nevertheless, one may question the barriers between various IHDIs, and the subsequent ranking.  Of course, the Pearson correlation   coefficient belongs to the [-1, +1] interval, but its exact value depends on the reliability of the FDI and GDP data. On one hand, one may test the data reliability through Benford's law, or through Zipf's law, and observe outliers, - which can thereafter be neglected if thought to be unreliable.  Another method, going beyond the Zip's power law, in particular allowing some better observation of extremes and anomalous deviations, is the extended rank-size law (Ausloos and Cerqueti, 2016)  when one or both ends of the distribution deviate from the ideal power law. Grouping all the IHDI, we test the rank-size law in  Section \ref{ranksizelaw}  for observing if any scattering of data might be ambiguous.

\section{%3.2 
Methodology.}

As outlined here above, there exists a plethora of literature on the existence (or lack thereof) of the relationship between FDI and growth. Complex regression models are drawn to establish the framework through which some result is gathered. This paper takes a different approach; the model intentionally seeks to limit assumptions unless necessary, essential or in the form of general statements. The purpose of this methodology, at its most fundamental is to establish an empirical core which is set at the most rudimentary empirical level. The observation of trends and the establishment of accuracy of these observations is the intention.   

The methodology is based on an adjusted Pearson correlation to allow for the introduction of time lags (The data is presented in time series format.).  The classical  ("equal time") Pearson 's  product moment correlation coefficient, used to measure the degree of linear dependence between two variables, is usually defined  through
% \begin{equation} 
% \rho = \frac{n \sum_i^n X_i Y_i \; -\; ( \sum_i^n  X_i)(\sum_i Y_i)}{\sqrt{n\sum_ i^n X_i^2- (\sum_i^n X_i)^2}\;  \sqrt{n\sum_ i^n Y_i^2- (\sum_i^n Y_i)^2} } 
% \label{eq:Pearsoncorr} \end{equation} 
 %where $n$ refers to the number of values,   
%$X$ and  $Y$ refer to the two datasets used,  i.e.,  here, to be GDP and FDI, respectively.  

%A lag is then introduced in the growth variable term,  holding the FDI data as  an independent variable ($Y_i$); the lags are introduced into the growth rate ($X_i$) in the form of ($X_{i + 1}$), ($X_{i + 2}$) and ($X_{i + 3}$). This allows for the comparison of FDI with the corresponding growth rates delayed by up to three years.   

\begin{equation}
\rho_{X_t,Y_t}=\frac{cov(X_t,Y_t)}{\sigma_{X_t} \sigma_{Y_t}}= \frac{n \sum_t^n X_t Y_t \; -\; ( \sum_t^n  X_t)(\sum_t Y_t)}{\sqrt{n\sum_ i^n X_t^2- (\sum_t^n X_t)^2}\;  \sqrt{n\sum_ i^n Y_t^2- (\sum_i^n Y_t^2)} } 
\end{equation}
where $cov(X_t,Y_t)$ is the covariance between $X_t$ and $Y_t$; $\sigma_{X_t}$ and $\sigma_{Y_t}$ is the standard deviation of the $X_t$ and $Y_t$ distribution respectively, and where $n$ refers to the number of values.
 The lagged serial correlation between $X_t$ and $Y_{t-k}$ of order $k$ is

\begin{equation}\label{eqk}
\rho_{X_t,Y_{t-k}}=\frac{cov(X_t,Y_{t-k})}{\sigma_{X_t}\sigma_{Y_{t-k}}}.
\end{equation}
 In the following $X_t$ corresponds to the $GDP_t$ and $Y_t$ to $FDI_t$. Hence equation \eqref{eqk} introduces lags in $FDI_{t-k}$ in the form $Y_{t-1},\:Y_{t-2}$ and $Y_{t-3}$.
This allows for the comparison of $GDP$ with the corresponding $FDI$ lagged by up to three years.

 The lag has been limited to three years,  as the data analysis showed an increasing lack of coherence  beyond this time value. Indeed, further lags were attempted but yielded no noteworthy results.  Given the nature of statistics as well as the emphasis of this study, -  that is the observation of trends, increasing this time frame provides a larger time frame in which the study showed this correlation is lost., as was found and discussed by Miskiewicz (2012).  
 Thus, "long range correlation results", being meaningless,   are not reported in this study.  
 It should be obvious at once that this "meaningful time lag" finding should have much implication on economic policies and theories.
 
\subsection{%4.1
 Results}
 The results are presented and discussed according to the clustering formation based on the IHDI country value.
\subsubsection{%4.2 
Very High IHDI: S1}
The first subset of data involving the "Very High" dataset as grouped by their IHDI is shown in Table \ref{Table1}. %4.1. 
The data  shows a significant skewness towards the later lags.  The data suggests a negative correlation between FDI and GDP growth.  However, Finland and  Sweden  are outliers,, presenting a mixed positive and negative correlation depending on the lag; both countries present a positive correlation coefficient for no lag  (Lag0) and for Lag3.  For Lag2, all  correlation coefficients are negative.

 \begin{table} \begin{center}
\begin{tabular}[t]{|c| c|c|c|c|c|c|}
  \hline
       S1  &       \multicolumn{4}{|c|}{   Pearson Correlation Coefficient}\\\hline 
 Country	& 	Lag0		&	Lag1		&	Lag2		&	Lag3	 		\\ \hline
Australia	&	-0.0891	&	-0.1057	&	-0.1658	&	-0.1544	\\
Austria	&	-0.0324	&	-0.0578	&	-0.2531	&	-0.1171	\\
Canada	&	-0.1116	&	-0.3343	&	-0.4100	&	-0.2259	\\
Denmark	&	0.0731	&	-0.0698	&	-0.2549	&	-0.1092	\\
France	&	-0.2658	&	-0.4128	&	-0.4663	&	-0.4376	\\
Finland	&	0.1477	&	-0.2007	&	-0.2216	&	0.0011	\\
Germany	&	-0.0492	&	-0.0832	&	-0.3272	&	-0.3945	\\
Netherlands	& -0.1127	&	-0.2612	&	-0.5156	&	-0.356	\\
Sweden	&	0.1827	&	-0.0798	&	-0.0671	&	0.0903	\\
Great Britain	& -0.0281	&	-0.2873	&	-0.2852	&	-0.2979	\\
Norway	&	-0.292	&	-0.5401	&	-0.4867	&	-0.5135	\\
Iceland	&	0.3606	&	0.2447	&	-0.0723	&	-0.4235	\\
Ireland	&	-0.051	&	-0.189	&	-0.3079	&	-0.3582	\\
\hline
\end{tabular}
 \caption{ %Table 4.1 :  
Time lag dependent  Pearson correlation  coefficients for  13 Very High Inequality-Adjusted Human Development Index (IHDI) countries } \label{Table1}
\end{center} \end{table}

\subsubsection{%4.4 
 High IHDI: S2}

The second data subset ,  "High" according to their  IHDI is analyzed in Table \ref{Table4} % 4.4  .  
The pattern which emerges  shows a significant skewness towards the later lags, but in a smaller degree than  seen in Table  \ref {Table1}. %4.1. 
for the "Very High". Although the dominant lag is again Lag2, in this case the second dominant lag is Lag1 followed by Lag3 and Lag0.  
 This data set too contains two outliers,  Argentina  and Cyprus, presenting a  positive correlation, at "no lag", in contrast to all other countries. Notice that Portugal  shows   a high  (negative) correlation at all levels, with a systematic tendency.

 \begin{table} \begin{center}
\begin{tabular}[t]{|c| c|c|c|c|c|c|}
  \hline
S2	        &       \multicolumn{4}{|c|}{   Pearson Correlation Coefficient}\\\hline 
  Country& Lag0	 	&	Lag1	&	Lag2	&	Lag3	 		\\ \hline
Argentina 	&	0.070	&	-0.0026	&	-0.1665	&	-0.2003	\\
Israel	&	-0.0814	&	-0.1962	&	-0.2664	&	-0.2507	\\
Spain	&	-0.1723	&	-0.3860	&	-0.4221	&	-0.3608	\\
Italy	&	-0.2451	&	-0.2707	&	-0.5857	&	-0.5027	\\
USA	&	-0.1703	&	-0.3730	&	-0.3959	&	-0.3118	\\
Portugal	&	-0.4676	&	-0.4155	&	-0.4101	&	-0.4005	\\
Greece	&	-0.1707	&	-0.3173	&	-0.4333	&	-0.5772	\\
Japan	&	-0.4620	&	-0.5979	&	-0.3580	&	-0.0681	\\
Malta	&	-0.2271	&	-0.3781	&	-0.1920	&	-0.2168	\\
Cyprus	&	0.0037	&	-0.0115	&	-0.0653	&	-0.3012	\\
Korea	&	-0.4883	&	-0.4884	&	-0.4548	&	-0.4838	\\
 \hline
\end{tabular}
 \caption{ %Table 4.5, - 
Time lag dependent Pearson correlation  coefficient values  for  11 High IHDI  countries} \label{Table4}
\end{center} \end{table}

\subsubsection{%4.5, - 
Medium IHDI: S3}

 The Pearson correlation  coefficient values  for  the medium IHDI  are shown in Table  \ref{Table6}. %4.8 
They show some skewness towards lower lags with no lag the dominant result, whatever the sign of the correlation,  followed by Lag2, Lag1 and  Lag3 respectively.    
The pertinent  findings about  the Very High and High datasets is mirrored here:  the correlation between FDI and Growth is no  longer predominantly negative but has shifted towards an  even split between positive and negative  correlations.  The positive correlation rather decreases with increasing lag.

 \begin{table} \begin{center}
\begin{tabular}[t]{|c| c|c|c|c|c|c|}
  \hline
 S3        &       \multicolumn{4}{|c|}{   Pearson Correlation Coefficient}\\\hline 
  Country & Lag0		&	Lag1	&	Lag2	&	Lag3	 		\\ \hline
Uruguay	&	0.2719	&	0.2009	&	0.1545	&	0.0954	\\
Sri Lanka	&	0.3837	&	0.2367	&	0.1429	&	0.1974	\\
%Trinidad \& Tobago	&	0.1617	&	0.0730	&	0.1136	&	0.0910	\\
Venezuela	&	0.1176	&	-0.0214	&	-0.3486	&	-0.1883	\\
Mexico	&	-0.2526	&	-0.3069	&	-0.2489	&	-0.1737	\\
Peru	&	0.3097	&	0.2389	&	0.1542	&	0.1571	\\
Mauritius	&	-0.0059	&	-0.0971	&	-0.1590	&	-0.1252	\\
Chile	&	-0.0805	&	-0.2090	&	-0.2509	&	-0.2252	\\
Turkey	&	0.0113	&	-0.1529	&	-0.1281	&	-0.0046	\\
\hline
\end{tabular}
 \caption{ %Table 4.5, - 
Time lag dependent Pearson correlation coefficient values for 8 Medium  IHDI  countries } \label{Table6}
\end{center} \end{table}

\subsubsection{%4.6, - 
Low IHDI: S4}

Lastly, this process is repeated for the Low HDI dataset. The data in Table \ref{Table9} %4.11 
skews towards the weaker lags with the absence of a lag again being  the dominant cases, followed by Lag1,  Lag2 and then  Lag3.  
 The pertinent  observation here concerning the correlation  shows  that  the Pearson coefficient is  predominantly positive with a few  negative correlations.

 \begin{table} \begin{center}
\begin{tabular}[t]{|c| c|c|c|c|c|c|}
  \hline
 S4        &       \multicolumn{4}{|c|}{   Pearson Correlation Coefficient}\\\hline 
  Country & Lag0		&	Lag1	&	Lag2	&	Lag3	 		\\ \hline
Philippines	&	0.2148	&	0.2372	&	0.1376	&	0.1110	\\			
%Colombia	&	0.0476	&	-0.0295	&	-0.0498	&	-0.0092	\\				
Paraguay	&	-0.1824	&	-0.1169	&	-0.0791	&	-0.1733	\\				
%Iran*	&	-0.0241	&	-0.0173	&	-0.0389	&	-0.0046	\\				
Iraq	&	-0.0876	&	-0.0362	&	-0.0849	&	-0.0626	\\				
Bolivia	&	0.2487	&	0.1702	&	0.1313	&	0.1387	\\				
%Morocco	&	-0.0640	&	-0.0529	&	-0.0348	&	-0.0185	\\				
SouthAfrica	&	0.0587	&	-0.0702	&	-0.0065	&	0.1224	\\			
Nigeria	&	0.1376	&	0.2380	&	0.2015	&	0.1699	\\				
Niger	&	0.2407	&	0.2615	&	0.2344	&	0.2181	\\				
ElSalvador	&	0.1095	&	0.0032	&	-0.0744	&	0.0461	\\			
India	&	0.2701	&	0.3192	&	0.2934	&	0.2574	\\				
Nepal	&	0.0722	&	0.0668	&	0.0241	&	0.0352	\\				
Ghana	&	0.4223	&	0.3874	&	0.3618	&	0.3487	\\				
\hline
\end{tabular}
 \caption{ %Table 4.5, - 
Time lag dependent  Pearson correlation coefficient values for 11  Low IHDI. } \label{Table9}
\end{center} \end{table}
	
\subsection{%4.7 - 
Discussion of findings}

In this subsection,  a closer look  can be taken at the findings of the numerical analysis.  

(1)	The data shows that as a country 's overall wealth increases, the effect of FDI  on stimulating growth becomes delayed. 

In wealthy "Very High" and "High" countries,  there is clearly a tendency for the FDI to show its influence upon the growth rate in a stronger way when considering  lags.  In the "Medium" and "Low" countries this influence is seen to be weaker  in the earlier stages, thus for no lags.

As such, conversely,   we can  conclude  that countries   with larger time delays  in their FDI/Growth relationship are those having higher GDP per capita.  

(2) It  can be also concluded that as a country overall wealth rises,  an increasing time lag becomes evident  in its growth  responsiveness from   FDI.  
This  can be  explained using macroeconomic theory.

Given the state of development in developing economies, FDI is used to increase the productive output of the nation. On the other  hand, investments made into developing countries  are often made in the form of "Green Field  Investment"\footnote{A "green field investment" is a type of foreign direct investment (FDI) where a parent company builds its operations in a foreign country from the ground up, in contrast to other methods of FDI, such as foreign acquisitions or buying controlling stakes in a foreign company. }.  This  policy takes advantage of the relatively lower costs of labour, materials, production and access to consumer markets than those  in a developed economy. This grants the parent firm a competitive advantage in the new market,  while also providing an entirely new audience to push their products on. It is known that Western Europe and North America accounted for 70\% of Green Field investment into Africa in 2014 (African Investment Report, 2015).  When this FDI inflow is invested in labour, materials and construction of infrastructure,    the FDI becomes reflected in the country's GDP. This explains why the FDI has a tendency to reflect itself in GDP (and thus growth) over the short run for developing economies.   
 Given the theoretical role of monopolistic power, this imperfect competition  of course creates barriers to entry for  domestic firms, whence harming domestic production in the long run, thereby perhaps explaining the short term causality which is observed.

 In developed economies, the opposite holds. Although Green Field Investment plays a part, it is not the driving factor. Although there are many reasons for FDI, inflows into developed economies occur for one or more of a few reasons, like
 \begin{itemize}
 \item 
 Stability: most FDI   into developed countries flows into established organizations. For investment purposes,  this provides a stable asset as part of a portfolio. Diversification in this manner allows for hedging of risk and provides access to established markets. 
\item
 Access to new technology: having access to new technology and organizational systems allows the parent firm to export and implement these technologies and techniques in their own  companies (elsewhere).
\end{itemize}
It is also widely accepted that business markets in developed economies tend to be saturated when compared to developing economies, -  with less  easy room for growth. In these, investments are not mostly inspired by maintaining and increasing competitiveness, -  in contrast with Green Field Investment  in developing economies. There is also some increased bureaucracy  (and/or corruption) which combined with the aforementioned causes,  increases the barriers  size to FDI entry.  Increased market penetration thus requires innovation and productive efficiency which intrinsically contribute towards the presence of a time lag, as here above demonstrated. 

 (3)	 
Developing countries have a predominantly positive correlation, whereas the wealthier countries have a predominantly negative FDI to Growth correlation.   This can be explained using macroeconomic theory:   although capital inflows into developing countries  quickly reflect   into their GDP, it takes relatively a longer  time for returns on investments to flow back to the developed nations. Thus,  within the limit of 3 year lags, this relationship is predominantly negative.

(4) The role of 	IHDI rankings.
We use  the IHDI in grouping the data sets; the ranking can be seen as a reasonably reliable approximation of economic progress. Given the consideration of inequality in income, consumption and wealth, the IHDI prevents resource-rich economic oligarchies from over performing. In this case a lower IHDI rank indicates a higher (better) position. 
 
Given the role of the IHDI in categorizing our data and its reflection of economic prosperity this result is reasonably expected.   The rank-size analysis  of the Pearson correlation coefficients, in the next section,  reflects such an apparently logical   {\it  but a priori}  grouping; it will  demonstrate the IHDI  influence as an independent parameter to be considered in further work and modeling.
 
\begin{figure}
 \centering
 \includegraphics  [height=20.1cm,width=13.1cm]{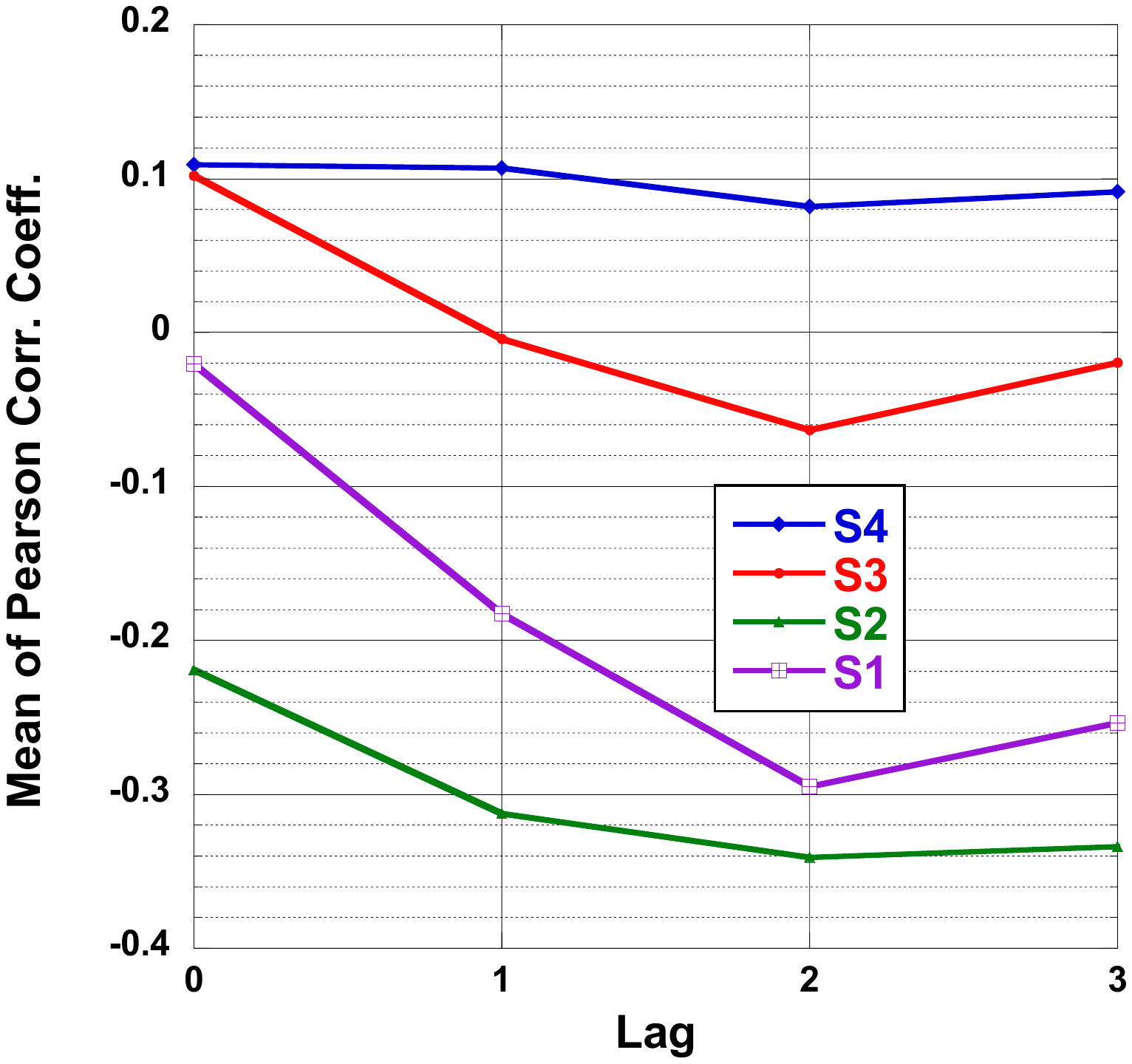} 
  \caption   { Evolution of the "averaged" Pearson correlation coefficient trend as a function of the time lags between FDI and GDP for the 4 clusters of countries. }
\label{fig:Plot43meansforlagsS1S2S3S4}
 \end{figure}

(5) The "relaxation time" or more positively the "memory effect" following FDI.  This  can be observed through Fig. \ref{fig:Plot43meansforlagsS1S2S3S4}. Due to various  expected differences between economies, one cannot expect a  finely defined evolution of the features through the Pearson correlation coefficient. However, an average "time lag  evolution" can be searched. In order to do so, we have averaged a linear fit  (the "trend") to each  correlation coefficient, in each sub-sample. The resulting break at the 2-3 year time interval is remarkable, pointing to the verification of our hypothesis on the possible observation of  the existence of such a lag effect  between FDI and GDP growth. The sign also adds value to the comment (4)

\subsection{%5. -
 Limitations}

Despite the findings, it must be noted that given the nature of the methodology adopted there remain several  possible  debates. One can note, and we admit it,   that the findings in no way imply causality. Purposefully,  we avoid Granger causality testing. We emphasize that although the results show that the variables FDI and GDP growth are correlated and  do confirm the presence of a lag, itself correlated with the state of a country 's  economic development, this does not absolutely establish direct causality. The Pearson correlation mainly measures the strength of a linear association, - not the cause.  For going beyond such a discussion, regression models  should be used.

The limitations of the interpreting power of the Pearson correlation is particularly relevant for the secondary observations found in the data; developed economies tended towards a negative Pearson coefficient between FDI and GDP growth whereas for developing nations gravitated towards a positive coefficient. In Table \ref{Table1}  
"Very High IHDI"  correlation coefficients are predominantly negative whereas in  Table \ref{Table9}   
 the sign of the "Low IHDI"  correlation coefficients is somewhat  evenly distributed.   
The Pearson correlation is a useful tool for the spotting of trends.    
 Economies however are incredibly complicated; the use of a linear method might be considered  too simplistic.  Nevertheless, the time lag effect seems indubitable, whence to be considered in causality tests. 
 
 Yet, one should think further on the time lag (Cerqueti et al.,  2018). Although this study significantly concludes on the existence of a finite size empirical GDP dependent time lag, usually about not more than 3 years,   
 its use remains debatable and retrospective.  
  Indeed, the   GDP data  is announced several months after its relevant data collection. Moreover,  the observations are not continuously  obtained, but  occur at discrete time intervals.  A finer approach, through trimester data for example, might be very valuable, surely  under presently globalization  conditions. 

  As such this perhaps represents an area for future study in order to establish relevance of the findings with regards to policy, ultimately seeking to aid in the development of economies.

\section{%6 - 
Rank-size Law}\label{ranksizelaw}

When the definition of data intervals can be debated,   in the context of best-fit procedures, the rank-size theory allows to
explore the presence of regularities among data and their  {\it a priori}  specified 
criterion-based ranking  (Jefferson, 1989; Vitanov and Ausloos, 2015).  Such regularities are
captured by a best-fit curve.  
In presence of an inflection point in a visually smooth data distribution, one could identify two
regimes in the ranked data, meaning that the values are clustered in
two families at   low and high ranks.  

A warning is in order:  it should be obvious that the ranking   of a country Pearson correlation coefficient  may change  from a  time lag  to another.   No doubt that the time dependence of the ranking  should be of interest as well, but  such a subject  is left for  dynamic evolution studies,  much  outside the present aims. One could also complement such an analysis through a
Kendall $\tau$ measure in order to observe the ranking consistency inside the IHDI groups or when moving from one group to another.

\begin{figure}
 \centering
 \includegraphics  [height=20.1cm,width=13.1cm]{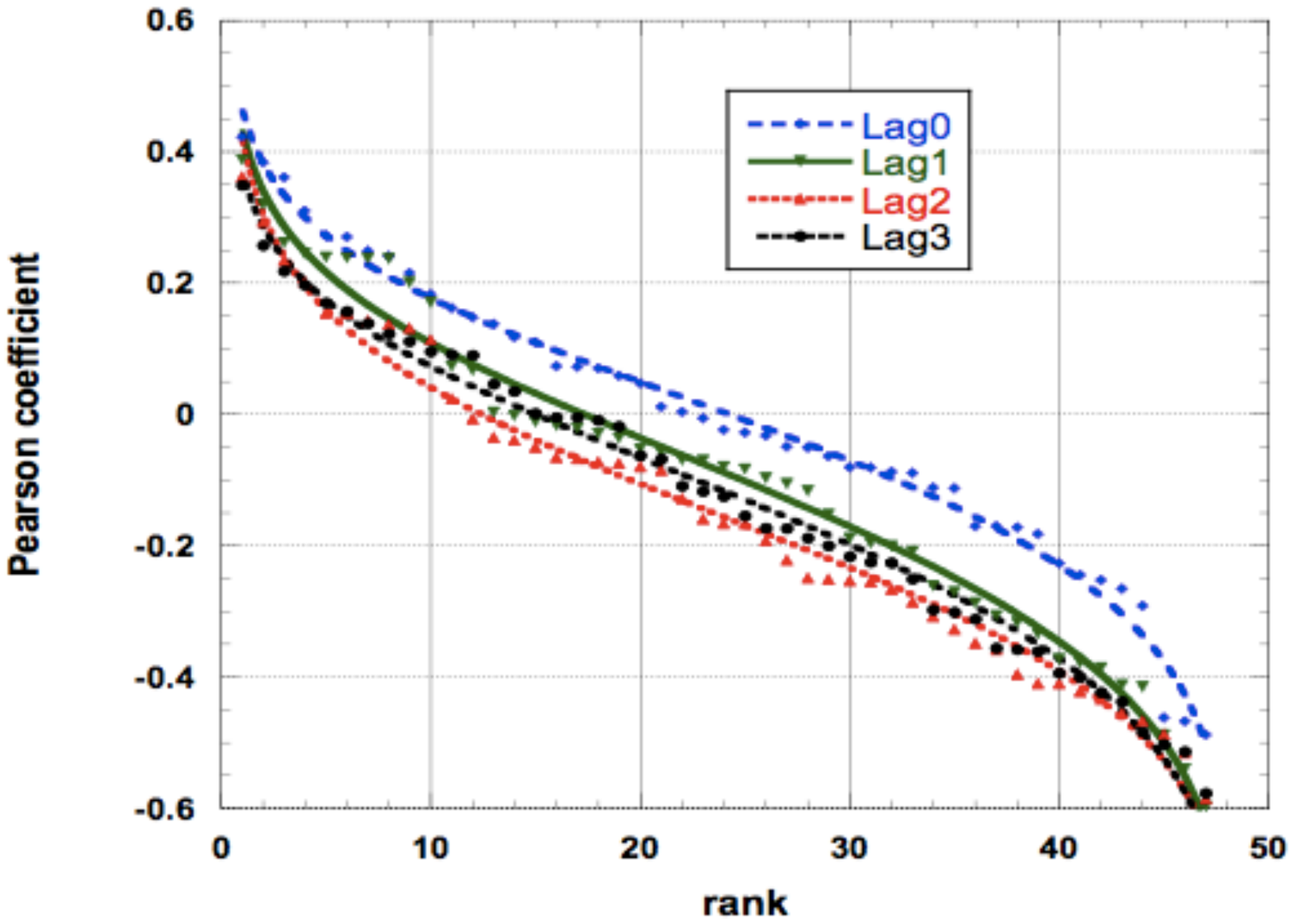} 
  \caption   { Rank-size law and its empirical fit, Eq.(\ref{Lav3}),  for the Pearson coefficient at different time lags between FDI and GDP}
\label{fig:Plot43PearsonLav3}
 \end{figure}

 \begin{table} \begin{center}
\begin{tabular}[t]{|c| c|c|c|c|c|c|}
  \hline
 Fit     &       \multicolumn{4}{|c|}{   Pearson correlation coefficient}\\
parameters  & Lag0		&	Lag1	&	Lag2	&	Lag3	 		\\ \hline
$m_1$	&	0.8657$\pm$0.0526	&	0.7365$\pm$0.0770&	0.9586$\pm$ 0.0964	&	0.7194$\pm$0.0639	\\
$m_2$	&	0.0773$\pm$0.0059&	0.0940$\pm$0.0097	&	0.1295$\pm$0.0093	&	0.0955$\pm$0.0082	\\
$m_3$&	0.2180$\pm$0.0089&	0.2733$\pm$0.0161&	0.2374$\pm$0.0156	&	0.2706$\pm$0.0137	\\ \hline
$R^2$& 0.9880 & 0.9771 & 0.9787 & 0.9832 \\
\hline
\end{tabular}
 \caption{ %Table 4.5, - 
Parameters of the rank-size law fit, Eq.(\ref{Lav3}),  to Pearson correlation coefficient values  at different time lags} \label{Tablernaksize}
\end{center} \end{table}
Therefore, a  rank-size rule fit was attempted with a   decreasing power law, with different exponents at low and high ranks, -  in order to obtain an inflection point near the center of the data range, i.e. with the analytical form  (Ausloos \&   Cerqueti, 2016)

\begin{equation}
y(r)\;=-1+m_1\;N^{-m_2}\;r^{-m_2}\;(N+1-r)^{m_3} 
 \label{Lav3} \end{equation} 
 where $r$ is the rank and $N= 43$. 
 The best  3-fit parameters have been so obtained:  $ m_1\in  [0.720;0.959] $;  $m_2 \in [0.077,0.130] $;  $m_3\in [0.218, 0.273]$: for a regression coefficient  $R^2 \sim [0.977,0.988]$, indicating a quite good agreement with Eq.(\ref{Lav3});  see Fig. \ref {fig:Plot43PearsonLav3} and Table \ref{Tablernaksize}. Notably, $m_3 >m_2$  is indicating a piling at low rank, i.e. for the "Low IHDI" countries (Ghana is always the $r=1$ country,  whatever the lag), - for which there is indubitably a positive Pearson correlation coefficient, thus for which FDI has markedly some influence on GDP. Notice, for example from Fig. \ref {fig:Plot43PearsonLav3}, that the number of positive correlations are for  roughly $r\le 20$, i.e. for the "Low IHDI" countries. The "Lag0" data is also quite different from those for finite lag, indicating, from this rank-size 	analysis,  that  to take into account time lags  when searching for FDI on GDP effects, is mandatory.

\section{%7 - 
Conclusions and perspectives}

This paper explores a novel area of research in the study of FDI.  We have  studied the relationship between Foreign Direct Investment and economic growth with regards to time lags. The study involved 43 countries over the period ranging from 1970-2015 with a total of 4278 observations.  Using the  Inequality-Adjusted Human Development Index   as a proxy for  an approximation of a country wealth, it has been found that the relationship between Foreign Direct Investment and the growth in Gross Domestic Product changes  sign depending on the countries approximate wealth. Wealthier countries as grouped in  "Very High" and "High" converge to an increased time lag for this relationship to exist:  FDI takes a longer time to influence economic growth. Developing countries as grouped in "Medium" and "Low"  present  a shorter time lag in this  efficiency relationship: FDI takes less time to influence economic growth.  This difference is explained by the driving factors behind FDI: in developing economies FDI is predominantly "market seeking",  whereas, in developed economies, it is  driven by demands for access and the diversification of investment. 

An interesting point seems to be the  (about 3 years) "memory effect" or "relaxation time effect",  between FDI inflow and growth output, suggesting some reflexion on  policies and strategies efficiencies.

\vskip0.5cm
{\bf Acknowledgments}

\vskip0.5cm
This paper is partially based upon AE's M.Sc. dissertation at ULSB under MA's supervision.

We gladly thank all reviewers for their comments.
 \newpage

 \clearpage  
{\bf References}
\vskip0.4cm
Abbes, S.M., M.B. Mostefa,  G.M.  Seghir,  \& G.Y. Zakarya. (2015).   Causal interactions between FDI, and economic growth: evidence from dynamic panel co-integration.   Procedia Economics and Finance,  23,  276-290. \\

Abernathy, J.L.,  Barnes, M.,  Stefaniak, C., \& Weisbarth, A. (2017).  An international perspective on audit report lag: A synthesis of the literature and opportunities for future research.  International Journal of Auditing,  21, 100--127.\\

Aga, A.A.K. (2014).  The impact of foreign direct investment on economic growth: A case study of Turkey 1980-2012.   Int. J. Econ. Finance,    6, 71-84.\\

Ahmed, E.M.  (2012).  Are the FDI inflows spillover effects on Malaysia s  economic growth input driven?   Economic Modelling,   29, 1498-504.\\

Aitken, B.J. \& A. Harrison. (1999).  Do Domestic Firms Benefit from Direct Foreign Investment? Evidence from Venezuela. American Economic Review,    89, 605-618.\\

Alfaro, L., A. Chanda, S. Kalemli-Ozcan, \& S. Sayek. (2004).  FDI and Economic Growth: The Role of Local Financial Markets. Journal of International Economics,    64, 113-134.\\

Ali, S.S., M.J. Anwer, H. Mustafa,  \& A.F. Sidiqqi. (2016).  Measuring the Dependence of Economic Growth (GDP) on Foreign Direct Investment, Labor Force, and Literacy Rate: The Case of 30 Selected Countries. Int. J. Adv. Multidiscip. Res,   3, 41--47.\\

 Almfraji, M.A. \& M.K. Almsafir. (2014).   Foreign direct investment and economic growth literature review from 1994 to 2012.  Procedia-Social and Behavioral Sciences,  129, 206-213.\\

Ausloos, M. \& R. Cerqueti. (2016).  A universal size-rank law. PLoS ONE,     2016, 0166011.\\

Ausloos, M. \& R.  Lambiotte.  (2007).  Clusters or networks of economies? A macroeconomic study through Gross Domestic Product. Physica A,    382, 16-21\\

Balasubramanayam V. N., M. Salisu, \& D. Sapsford. (1996).  Foreign Direct Investment and Growth in EP and IS Countries,  The Economic Journal,    106, 92-105.\\

Balasubramanyam, V.N.,  M. Salisu,  \& D. Sapsford. (1999).  Foreign Direct Investment as an engine of Growth.   The Journal of International Trade \& Economic Development,   8, 27-40.14 \\

Barrell, R., A. Nahhas,  \& J. Hunte.  (2017). Exchange Rates and Bilateral FDI: Gravity models of Bilateral FDI in High Income Economies. Economics and Finance Working Paper Series, Brunel University London, (17-07). \\

Barrell, R. \& A. Nahhas, A. (2018). Economic integration and bilateral FDI stocks: the impacts of NAFTA and the EU.  LSE Research Online Documents on Economics 90372, London School of Economics and Political Science, LSE Library.\\

Belloumi,  M.  (2014).  The relationship between trade, FDI and economic growth in Tunisia: an application of the autoregressive distributed lag model.   Economic Systems,   38, 269-287.\\

Bengoa, M., \& B. Sanchez-Robles. (2003).  FDI, Economic Freedom, and Growth: New Evidence from Latin America. European Journal of Political Economy,    19, 529-545.\\

Blomstr{\"om, M. \& A. Kokko. (1996).  Multinational Corporations and Spillovers.  Journal of Economic Surveys,    12, 247-277.\\

Blomstr\"om, M. \& A. Kokko. (2003).  The Economics of Foreign Direct Investment Incentives. NBER Working paper 9489. \\

  Blonigen, B.A., R. B. Davies, G. R. Waddell,  \& H.T. Naughton. (2007).  FDI in space: Spatial autoregressive relationships in foreign direct investment. European Economic Review,  51,  1303-1325. \\

Borensztein, E., J. De Gregorio, \& J-W. Lee. (1998).  How Does Foreign Direct Investment Affect Economic Growth?   Journal of International Economics,    45, 115-135.\\

Bos, H.C., M. Sanders, \&  C. Secch. (1974).  Private foreign investment in developing countries. (Dordrecht, Holland: D. Reidel).\\

Brambilla, I.,  G. Hale,   \& C. Long.  (2009).  Foreign direct investment and the incentives to innovate and imitate. The Scandinavian Journal of Economics,    111,   835--861. \\

Braunstein, E. \& G. Epstein.  (2002).   Bargaining Power and Foreign Direct Investment in China: Can 1.3 Billion Consumers Tame the Multinationals?, Working paper (Amherst, MA: University of Massachusetts).  \\

Bushman, R.M. \& C.D. Williams. (2015).  Delayed expected loss recognition and the risk profile of banks.  Journal of Accounting Research,  53, 511--553.\\

Carkovic, M. \& R. Levine.  (2005).   Does Foreign Direct Investment Accelerate Economic Growth?  in  T.H.  Moran, E. M. Grahan, \& M. Blomstr\"om (eds.),  Does Foreign Direct Investment Promote Development?,  Washington D.C.: Institute for International Economic, 195-220.\\

Caves, R.E.  (1971).  International corporations: the industrial economics of foreign investment. Economica,    38, 1-27. \\
	
Caves, R.  (1974).  Multinational Firms, Competition and Productivity in the Host Country. Economica,    41, 176-193.  \\

Cerqueti, R., L. Fenga,  \& M. Ventura,  (2018). Does the US exercise contagion on Italy? A theoretical model and empirical evidence. Physica A: Statistical Mechanics and its Applications, 499, 436-442.\\

Chowdhury, A. \&  G. Mavrotas.  (2005).   FDI and growth: a causal relationship, Research Paper, UNU-WIDER, United Nations University (UNU)\\

Choe, J.I.  (2003).  Do foreign direct investment and gross domestic investment promote economic growth?   Review of Development Economics,    7,  44--57.\\
  
De Gregorio, J. (2005).  The role of foreign direct investment and natural resources in economic development.  in : Graham E.M. (eds) Multinationals and Foreign Investment in Economic Development. International Economic Association Series. Palgrave Macmillan, London, 179-197.
%Working paper No 196. Central Bank of Chile, Santiago\\
\\

De Mello, L.R.   (1999).  Foreign Direct Investment-Led Growth: Evidence from Time Series and Panel Data. Oxford Economic Papers,    51,  133-151.\\

Durham,   J.B. (2004),  Absorptive Capacity and the Effects of Foreign Direct Investment and Equity Foreign Portfolio Investment on Economic Growth,  European Economic Review,    48, 285-306.\\

Ericsson, J. \& M. Irandoust.  (2001).  On the causality between foreign direct investment and output: a comparative study,  The International Trade Journal,    15, 1-26.\\

Germidis, D. (1977).  Transfer of technology by multinational corporations.  Paris. Development Centre of Organization for Economic Cooperation and Development.\\

G\"org, H. \& D. Greenaway. (2002).  Much ado about nothing? Do domestic firms really benefit from foreign direct investment?. CEPR Discussion paper DP3485.\\

   Griffin, R. W. \&  M.W. Pustay. (2007).  International Business: A Managerial Perspective  (5th ed.). (New Jersey: Pearson/Prentice Hall). \\
 
Haddad, M. \& A. Harrison. (1993).  Are There Positive Spillovers from Direct Foreign Investment?: Evidence from panel data for Morocco. Journal of Development Economics,    42, 51-74.\\

Hale, G. \& C. Long.  (2011).  Did foreign direct investment put an upward pressure on wages in China?.  IMF Economic Review,   59, 404--430. \\

Hansen, H. \&  J. Rand.  (2006).   On the causal links between FDI and growth in developing countries.   The World Economy,    29, 21-41.\\

Harttgen, K., \& S. Klasen. (2012). A household-based human development index. World Development, 40(5), 878-899.\\

Hermes, N., \& R. Lensink.  (2003).  Foreign direct investment, financial development and economic growth.    Journal of Development Studies,   40,  142-163.\\

Herzer, D., S. Klasen, \& F.N. Lehmann.  (2008).  In search of FDI-led growth in developing countries: the way forward.   Economic Modelling,    25, 793-810.\\

Hoekman, B. \&  B.S. Javorcik. (2004).   Policies facilitating firm adjustment to globalization.   Oxford Review of Economic Policy,  20,  457-473. \\

Huang, Y. (1998).  FDI in China: an Asian perspective. Singapore: Institute of Southeast Asian Studies.\\

Huang, Y. (2003).  One country, two systems: foreign-invested enterprises and domestic firms in China.   China Economic Review,   14, 4040-4016.\\

 Hymer, S.  (1960).  On Multinational Corporations and Foreign Direct Investment , in  J. H. Dunning (ed.),  The Theory of Transnational Corporations. London: Routledge for the United Nations.\\

 Irandoust, J. E. M. (2001).  On the causality between foreign direct investment and output: a comparative study,   The International Trade Journal,  15, 1-26.\\
 
Javorcik, B. S. \& K. Saggi.  (2010).  Technological asymmetry among foreign investors and mode of entry,  Economic Inquiry,  48, 415--433.\\

Jefferson, M. (1989).  Why geography? The law of primate city.      Geographical  Review,    79,  226- 232.\\

Kuo, K. C.,   C. Y. Chang, M. H.  Chen,  \& W. Y. Chen. (2003).  In search of causal relationship between FDI, GDP, and energy consumption-Evidence from China.   Advanced Materials Research,    524, 3388-3391. \\
  
Li, X.  \& X. Liu. (2005).  Foreign direct investment and economic growth: an increasingly endogenous relationship.   World development,   33, 393-407.\\

 Li, J. C.   \&  D.C. Mei.  (2013).  The influences of delay time on the stability of a market model with stochastic volatility.   Physica A,      392, 763-772 \\

Liang, X. S. (2014).  Unraveling the cause-effect relation between time series.   Phys. Rev. E,   90, 052150-1-11. \\

Liang, X. S. (2016).  Information flow and causality as rigorous notions ab initio.   Phys. Rev. E,    94, 052201-1-28.\\

Lipsey, R.  E.  (2002).  Home and Host Country Effects of FDI, NBER Working paper 9293; ibid.  (2002).   Foreign direct investment, growth, and competitiveness in developing countries.   The global competitiveness report,  2003, 295--305, \\

Lipsey, R.E. \& F.  Sj\"oholm. 2006).  Foreign Firms and Indonesian Manufacturing Wages: An Analysis with Panel Data.   Economic Development and Cultural Change,    55, 201-221\\

 Louzi, B.M., \& A. Abadi.  (2011).  The foreign direct investment on economic growth in Jordan.     International Journal of Research \& Reviews in Applied Sciences,    8, 253-258.\\
	
Mah, J.S. (2010).  Foreign Direct Investment Inflows and Economic Growth of China.   Journal of Policy Modeling,    32, 155-158.\\
	
Mansfield, E. \& A. Romeo, (1980).  Technology transfers to overseas subsidiaries by US-based firms.   Quarterly Journal of Economics,    95, 737-750.\\
	
Marin, D. \& M. Schnitzer.  (2011).  When is FDI a capital flow?   European Economic Review,  55, 845-861.\\
	
Mazenda, A.  (2014).  The effect of foreign direct investment on economic growth: evidence from South Africa.   Mediterranean Journal of Social Sciences,    5, 95-108.\\
	
Mencinger, J.   (2003).  Does foreign direct investment always enhance economic growth?. Kyklos,    56, 491--508. \\

Miskiewicz, J. (2012).   Analysis of time series correlation. The choice of distance metrics and network structure.   Acta Physica Polonica A,   121,  B89-B94.\\
	
Nair-Reichert, U. \& D. Weinhold. (2001).  Causality Tests for Cross-Country Panels: a New Look at FDI and Economic Growth in Developing Countries.   Oxford Bulletin of Economics and Statistics,   63, 153--171.\\

Rappaport, J. (2000).  How Does Openness to Capital Flows Affect Growth?, Mimeo, Federal Reserve Bank of Kansas City,  RWP 00-11.  \\

Romer, P. (1993).   Idea gaps and object gaps in economic development.   Journal of Monetary Economics,   32, 543-573.\\
	
Saltz, S. (1992).  The Negative Correlation Between Foreign Direct Investment and Economic Growth in the Third World: Theory and Evidence. Rivista Internazionale di Scienze Economiche e Commerciali,    39,  617-633.\\
	
Strat, V. A. (2014). What happened with the attractiveness of the Romanian counties for FDI during the period 2001-2012?. Journal
of Applied Quantitative Methods,   9, 22-37. \\

Temiz, D., \& A.  G\" okmen.   (2014).  FDI inflow as an international business operation by MNCs and economic growth: An empirical study on Turkey.   International Business Review,    23, 145-154.\\

 Tu, Y. \& X. Tan. (2012).  Technology spillovers of FDI in ASEAN sourcing from local and abroad.   China Finance Review International,  2, 78-94.\\

UNDP. (2006). Human Development Report 2006. Beyond Scarcity: Power, Poverty and the Global Water Crisis.UNDP, New York.\\

Vitanov, N.K., \& M. Ausloos.  (2015). Test of two hypotheses explaining the size of populations in a system of cities. Journal of Applied Statistics, 42, 2686-2693.

Wang, D.T.,  F. F. Gu,   D.K. Tse,  \&   C. K. B. Yim, (2013).  When does FDI matter? The roles of local institutions and ethnic origins of FDI.
  International Business Review,  22, 450--465. \\
	
World Bank (2016) World Bank Open Data. 
Available at: 

http://data.worldbank.org/  
\\

Zhang, K.H. (2001).  Does foreign direct investment promote economic growth? Evidence from East Asia and Latin America.    Contemporary economic policy,    19, 175--185.

	\end{document}